\documentclass{elsart} 
 
\usepackage{amsmath} 
\usepackage{mathrsfs} 
\usepackage[T1]{fontenc} 
 
\begin{document} 
 
\runauthor{Brodin, Stenflo, Anderson, Lisak, Marklund and Johannisson}   
 
\begin{frontmatter} 
 
\title{Light bullets and optical collapse in vacuum} 
 
\author[Umea]{Gert Brodin}, 
\author[Umea]{Lennart Stenflo}, 
\author[Goteborg]{Dan Anderson}, 
\author[Goteborg]{Mietek Lisak}, 
\author[Goteborg]{Mattias Marklund} and 
\author[Goteborg]{Pontus Johannisson}  
 
\address[Umea]{Department of Plasma Physics, Ume{\aa} University, 
SE--901 87 Ume{\aa}, Sweden}  
\address[Goteborg]{Department of Electromagnetics, Chalmers University of  
Technology, SE--412 96 G\"oteborg, Sweden}  
 
\date{\today} 
 
\begin{abstract} 
In quantum electrodynamics, photon--photon scattering can be the  
result of the exchange of virtual electron--positron 
pairs. Effectively,  
this gives rise to self-interaction terms in Maxwell's equations, 
similar to  
the nonlinearities due to polarization in nonlinear optics. These 
self-interaction terms vanish in the limit of parallel propagating 
waves.  
However if the modes generated in bounded regions are used, there will 
be a  
non-zero total effect. We show that stationary two-dimensional light 
bullets can form  
in guiding structures, due to the balancing effect of quantum 
electrodynamical vacuum nonlinearities on dispersion and diffraction. 
These light bullets are unstable and exhibit the possibility 
of self-focusing collapse. The consequences of our results are also 
discussed. \\[3mm] 
\noindent PACS numbers: 12.20.Ds, 42.50.Vk, 42.65.T 
\end{abstract} 
 
\end{frontmatter} 
 
%\maketitle 

%%%%%%%%%%%%%%%%%%%%%%%%%%%%%%%%%%%%  
\section{Introduction} 
%%%%%%%%%%%%%%%%%%%%%%%%%%%%%%%%%%%%  
 
According to quantum electrodynamics (QED), the non-classical 
phenomenon of   
photon--photon scattering can take place due to the exchange of 
virtual   
electron--positron pairs. This is a second order effect (in terms of  
the fine structure   
constant $\alpha\equiv e^{2}/4\pi \varepsilon _{0}\hbar c\approx 
1/137$), which in standard   
notation can be formulated in terms of the Euler--Heisenberg 
Lagrangian density   
\cite{Heisenberg-Euler,Schwinger}  
\begin{equation} 
{\mathscr{L}}_{{\rm EH}}=\varepsilon _{0}{\mathscr{F}}+ 
\xi (4{\mathscr{F}}^{2}+7{\mathscr{G}}^{2}) ,  \label{eq:Lagrangian} 
\end{equation} 
where $\xi \equiv 2\alpha ^{2}\varepsilon 
_{0}^{2}\hbar^{3}/45m_{e}^{4}c^{5}$,   
${\mathscr{F}}\equiv \frac{1}{2}(E^{2}-c^{2}B^{2})$,  
${\mathscr{G}}\equiv c{\bf E}\cdot {\bf B}$, $e$ is the electron 
charge,   
$c$ the velocity of light, $2\pi \hbar $ the Planck constant and 
$m_{e}$ the   
electron mass. The latter terms in expression (\ref{eq:Lagrangian}) 
represent the effects of vacuum   
polarization and magnetization. We note that 
${\mathscr{F}} = {\mathscr{G}}=0$ in the limit   
of parallel propagating waves. It is therefore necessary to use other 
waves in order  
to obtain an effect from the QED corrections. Furthermore, it has 
recently been shown that QED   
vacuum nonlinearities can lead to self-focusing of laser beams  
\cite{segev}, and that such effects might be experimentally viable 
within a decade.   
In Ref.\ \cite{PRL-article} it was found that guiding structures can 
be useful for  
the purpose of detecting QED effects, and a suggestion was presented 
for an   
experimental setup that gives a measurable signal.  
 
In the present work we thus consider   
waves guided by two parallel conducting planes, in order to study the 
possibility of  
using nonlinear QED effects to balance diffraction and dispersion in 
directions   
parallel to the planes. We find that our system is governed by a  
$1 + 2$ dimensional nonlinear Schr\"odinger equation (NLSE).  
Using approximate variational methods, which have proven to be useful   
in similar situations (see e.g.\ 
\cite{Desaix-Anderson-Lisak,Anderson-Cattani-Lisak}), we   
find that the nonlinearities can counteract diffraction and dispersion  
to form 2-dimensional   
light bullet solutions, i.e., solitary solutions to the NLSE which 
preserve the envelopes of the fields. 
The formation of light bullets was in other contexts (optical media)
first studied by Silberberg \cite{Silberberg}.

%%%%%%%%%%%%%%%%%%%%%%%%%%%%%%%%%%%%%%%%%%%  
\section{Light bullet solutions}\label{sec:two} 
%%%%%%%%%%%%%%%%%%%%%%%%%%%%%%%%%%%%%%%%%%%   
 
In a medium with polarization ${\bf P}$ and magnetization ${\bf M}$,  
the general wave equations for ${\bf E}$ and ${\bf B}$ are 
\cite{PRL-article}  
\begin{equation} 
\frac{1}{c^2}\frac{\partial^2{\bf E}}{\partial t^2} - \nabla^2{\bf E}   
=-\mu_0\left[ \frac{\partial^2{\bf P}}{\partial t^2} + 
c^2\nabla(\nabla\cdot{\bf P}) +   
\frac{\partial}{\partial t}(\nabla\times{\bf M)} \right]  
\label{WaveE}  
\end{equation} 
and  
\begin{equation} 
\frac{1}{c^2}\frac{\partial^2{\bf B}}{\partial t^2} - \nabla^2{\bf B} 
=   
\mu_0\left[ \nabla\times(\nabla\times{\bf M}) +  
\frac{\partial}{\partial t}(\nabla\times{\bf P)} \right]  . 
\label{WaveB}  
\end{equation} 
Furthermore, the effective polarization and magnetization in vacuum 
due to   
photon--photon scattering induced by the exchange of virtual 
electron--positron   
pairs are given by (see, e.g., Ref.\ \cite{segev})  
\[ 
{\bf P} = 2\xi\left[ 2(E^2 - c^2B^2){\bf E} + 7c^2({\bf E\cdot B}){\bf  
B} \right]   
\] 
and  
\[ 
{\bf M} = -2c^2\xi\left[ 2(E^2 - c^2B^2){\bf B} + 7({\bf E\cdot 
B}){\bf E} \right]  .   
\] 
We consider propagation between two parallel conducting planes with  
spacing $x_0$ (i.e., the region $0\leq x\leq x_0$ is vacuum surrounded  
by the plates that, as a starting point, are assumed to be perfectly 
conducting).   
We assume that only one ${\rm TE}_{n0}$ mode ($n = 1, 2, ...$) is 
present. To linear order, this gives the fields  
\begin{subequations} 
\label{eq:pump10} 
\begin{eqnarray} 
B_{z} \!&=&\! \frac{\pi }{x_{0}}\tilde{A}\, 
\cos \left( \frac{n\pi x}{x_{0}}\right) \exp [{\rm i}(kz-\omega 
t)]+{\rm c.c.} ,   
\\ 
B_{x} \!&=&\! -{\rm i}k\tilde{A}\,\sin \left( \frac{n\pi x}{x_{0}}\right)   
\exp [{\rm i}(kz-\omega t)]+{\rm c.c.} ,  
\\ 
E_{y} \!&=&\! {\rm i}\omega \tilde{A}\sin \left( \frac{n\pi 
x}{x_{0}}\right)   
\exp[{\rm i}(kz-\omega t)]+{\rm c.c.} , 
\end{eqnarray} 
\end{subequations} 
together with $0\approx \omega ^{2}/c^{2}-k^{2}-n^2\pi 
^{2}/x_{0}^{2}$. 
Here c.c.\ stands for complex conjugate. 

It should be noted that the beams propagate with a 90 degree angle
with respect to each other for the selfconsistent solutions obtained
in a similar situation in Ref.\ \cite{segev}. This would correspond to
the relation $k=n \pi/x_0$ in our case. However, we have no 
such limitation in our derivation. The result of a 90 degree beam
angle of Ref.\ \cite{segev} comes from requiring an explicit symmetry
between the wave equations for $\mathbf{E}$ and $\mathbf{B}$.  A
similar requirement is not needed in our case, since the
$\mathrm{TE}_{n0}$-modes introduced by the conducting planes do not
have a complete electric-magnetic symmetry as, for example, the
electric fields vanish close to the planes. What one still might worry 
about is whether or not both Faraday's and Ampere's law, which are
equivalent 
to the two wave equations for $\mathbf{E}$ and $\mathbf{B}$, 
are properly solved in our case. However, the equation that will be
derived for the vector potential is equivalent to Ampere's 
law, which is equivalent to the wave equation for the electric field,
and using the vector potential we therefore automatically solve Faraday's
law. It thus follows that our solution for the vector potential
corresponds to electromagnetic fields that are consistent with the two
wave equations for $\mathbf{E}$ and $\mathbf{B}$. 

We have expressed the 
fields in   
terms of the vector potential amplitude $\tilde{A}$, given by ${\bf 
A}=(0,A,0)$,   
where  
\begin{equation} 
A=\tilde{A}\sin \left( \frac{n\pi x}{x_{0}}\right)  
\exp [{\rm i}(kz-\omega t)]+{\rm c.c.} 
\end{equation} 
using the radiation gauge ($\phi =0$). A nonlinear dispersion relation  
can be derived   
by inserting the linear expression for the fields into the right hand  
side of Eq.\ (\ref{WaveB}), taking the  
$z$-component of the resulting equation, and  
separating into orthogonal trigonometric functions.  
From that equation the coefficients in the NLSE can be found. However,  
here we will present an alternative derivation, starting directly from 
the Euler-Heisenberg Lagrangian (\ref{eq:Lagrangian}), which is more  
elegant and gives the same result. Naturally we must first  
express $\mathscr{L}_{{\rm EH}}$ in terms of the electromagnetic  
potential, since it is the basis for the original variational 
principle.   
We take $\tilde{A} = \tilde{A}(t,z)$, and assume $\tilde{A}$ to be 
weakly   
modulated ($|\partial\tilde{A}/\partial t| \ll |\omega\tilde{A}|$, 
$|\partial\tilde{A}/\partial z| \ll  |k\tilde{A}|)$, but we omit   
the slow $y$-dependence, since such a dependence makes the derivation 
technically   
more complicated without adding extra understanding.  
A diffraction term in the $y$-direction is lost by this procedure, but  
this effect is trivially added afterwards.  
To lowest order, we thus omit the nonlinear terms and the   
slow derivatives in $\mathscr{L}$. Noting  
that ${\mathscr{F}}=(1/2)[(\partial A/\partial t)^{2}-c^{2}(\nabla 
A)^{2}]$ and   
${\mathscr{G}}=0$, we find that the non-oscillating part (i.e.\ the 
part that does   
not vanish after integration) is  
\begin{equation} 
{\mathscr{L}}_{0}/\varepsilon _{0}\equiv {\mathscr{F}}= 
\frac{1}{2}(\omega^{2}-k^{2}c^{2})|\tilde{A}|^{2}\sin ^{2}\left( 
\frac{n\pi x}{x_{0}}\right)   
-\frac{n^{2}\pi^{2}c^{2}}{2x_{0}^{2}}|\tilde{A}|^{2} 
\cos^{2}\left( \frac{n\pi x}{x_{0}}\right) . 
\end{equation} 
After performing the integration over the region between the plates we 
find that the   
lowest order part of the action vanishes identically as the dispersion 
relation is   
here considered to be satisfied. Going to the next order of 
approximation in the Lagrangian   
we include first order slow derivatives, which  
yields ${\mathscr{L}}_{1}/\varepsilon _{0}= 
{\mathscr{L}}_{0}/\varepsilon _{0}+{\rm i} 
\omega \lbrack (\partial A/\partial t)A^{\ast }-(\partial A^{\ast  
}/\partial t)A]-  
{\rm i}kc^{2}[(\partial A/\partial z)A^{\ast }- 
(\partial A^{\ast }/\partial z)A]$.  
After variation this leads to an equation where the envelope moves 
with the group velocity.   
The next order and final approximation includes second order slow 
derivatives and   
the ${\mathscr{F}}^{2}$ term in the Lagrangian. After performing the 
$x$-integration,   
dropping ${\mathscr{L}}_{0}$ (since the action is identically zero 
because of the dispersion   
relation) and eliminating the second order slow time-derivatives  
using $\partial^{2}/\partial t^{2}\approx v_{g}^{2}\partial 
^{2}/\partial z^{2}$, the final   
expression for the Lagrangian is  
\begin{eqnarray} 
\langle {\mathscr{L}}_{2}\rangle  
&=& {\rm i}\omega \varepsilon _{0}\left( \frac{\partial 
\tilde{A}}{\partial t}\tilde{A}^{\ast }-  
\frac{\partial \tilde{A}^{\ast }}{\partial t}\tilde{A}\right) - 
{\rm i}kc^{2}\varepsilon _{0}\left( \frac{\partial \tilde{A}}{\partial  
z}\tilde{A}^{\ast }-  
\frac{\partial \tilde{A}^{\ast }}{\partial z}\tilde{A}\right) 
\nonumber \\ 
 && \qquad \qquad \qquad \qquad +(c^{2}-  
v_{g}^{2})\varepsilon_{0}\left| \frac{\partial \tilde{A}}{\partial 
z}\right| ^{2}+  
\frac{3n^{4}c^{4}\pi ^{4}\xi }{x_{0}^{4}}|\tilde{A}|^{4} , 
\end{eqnarray} 
where $\langle \rangle $ stands for integration over 
$x$. Variation with respect to  $\tilde{A}^*$ leads to  
\begin{equation} 
  2{\rm i}\omega \left( \frac{\partial }{\partial t}+ 
v_{g}\frac{\partial }{\partial z}\right)A    
+ c^{2}\frac{\partial ^{2}A}{\partial y^{2}}+ 
\omega v_{g}^{\prime }\frac{\partial ^{2}A}{\partial z^{2}} 
+ \frac{6n^{4}c^{4}\pi^{4}\xi }{\varepsilon _{0}x_{0}^{4}}|A|^{2}A  
= 0  ,   
\label{nls}  
\end{equation} 
where $v_{g}$ is the group velocity and $v_{g}^{\prime }$ the group 
dispersion that   
follows from the linear dispersion relation. We have also added the 
diffraction term   
in the $y$-direction corresponding to the full amplitude  
dependence $A=\tilde{A}(t,y,z)$. Changing to a system moving with the 
group   
velocity while rescaling the coordinates and the amplitude according 
to  
\begin{equation} 
  \begin{array}{ll} 
    \tau  = \omega t/2\ , &  \upsilon =\,\omega y/c\ ,  \\[2mm] 
    \xi = \sqrt{\omega/v_{g}^{\prime }}\,(z-v_{g}t)\ , & a= 
       \displaystyle{\sqrt{\frac{6n^{4}c^{4}\pi ^{4}\xi }{\omega 
    \varepsilon_{0}x_{0}^{4}}}\,A}  ,   
  \end{array}  
\end{equation} 
we obtain  
\begin{equation} 
{\rm i}\frac{\partial a}{\partial \tau }+\frac{\partial 
^{2}a}{\partial\upsilon ^{2}}+  
\frac{\partial ^{2}a}{\partial \xi ^{2}}+ 
|a|^{2}a=0 .  
\label{nls2} 
\end{equation} 
which corresponds to the rescaled Lagrangian density  
\begin{equation} 
{\mathscr{L}}=\frac{{\rm i}}{2}\left( a^{\ast }\frac{\partial 
a}{\partial\tau }-  
a\frac{\partial a^{\ast }}{\partial \tau }\right) - 
\left| \frac{\partial a}{\partial \upsilon }\right| ^{2}- 
\left| \frac{\partial a}{\partial \xi }\right| ^{2}+ 
\frac{1}{2}|a|^{4} ,  \label{Lagrangian} 
\end{equation} 
where the diffraction term in the $y$-direction is also included.  
 
From now on we will look   
for cylindrically symmetric solutions of Eq.\ (\ref{nls2}), i.e., 
$a=a(t,\rho )$,   
where $\rho ^{2}=\upsilon ^{2}+\xi ^{2}$. Equation (\ref{nls2}) can 
then be written   
\begin{equation} 
{\rm i}\frac{\partial a}{\partial \tau }+ 
\frac{1}{\rho }\frac{\partial }{\partial \rho } 
\left( \rho \frac{\partial a}{\partial \rho }\right) 
+|a|^{2}a=0 ,  \label{nls3} 
\end{equation} 
while the Lagrangian density (\ref{Lagrangian}) takes the form  
\begin{equation} 
{\mathscr{L}}=\frac{{\rm i}}{2}\left( a^{\ast }\frac{\partial 
a}{\partial\tau }-  
a\frac{\partial a^{\ast }}{\partial \tau }\right) - 
\left| \frac{\partial a}{\partial \rho }\right| ^{2}+ 
\frac{1}{2}|a|^{4} ,  \label{Lagrangian2} 
\end{equation} 
with the action given by ${\mathscr{A}} =\int 
{\mathscr{L}}\rho\,d\rho\,{d\tau}$.   
Equation (\ref{nls3}) is a 2-dimensional radially symmetric NLSE.   
Although exact solutions of this equation are not available, numerical 
and approximate   
analysis give a clear picture of the solutions. In particular, it has 
been shown    
that Eq.\ (\ref{nls3}) allows a stationary solution where the 
diffractive/dispersive spreading   
of the pulse in the coordinate $\rho$ is balanced by the focusing 
effect created by the   
nonlinearity. An accurate analytical approximation of the dynamics of 
the pulse-like solutions   
of Eq.\ (\ref{nls3}) can be obtained using direct variational methods 
involving the   
Lagrangian ${\mathscr{L}}$ given by expression (\ref{Lagrangian2}) and  
subsequent Rayleigh--Ritz optimization based on suitably chosen trial  
functions (see e.g.\ \cite{Anderson-Cattani-Lisak} and references 
therein).   
A convenient trial function for the present problem is  
\cite{Desaix-Anderson-Lisak}  
\begin{equation} 
a_T(\tau, \rho) =  
F(\tau)\,{\rm sech}\!\left[ \frac{\rho}{f(\tau)}  
\right]\exp\left[ {\rm i} b(\tau)\rho^2 \right] 
\end{equation} 
which involves a complex amplitude $F(\tau)$, a pulse width $f(\tau)$, 
and a   
quadratic phase function modeling the phase front curvature.  
Inserting this ansatz into the variational integral and integrating 
over $\rho$, a reduced   
variational problem is obtained for the parameter  
functions, $F(\tau)$, $F^*(\tau)$, $f(\tau)$ and $b(\tau)$. The 
subsequent   
reduced Euler--Lagrange equations can be rearranged to give $F$, $F^*$  
and $b $ as   
explicit functions of the width $f(\tau)$, which satisfies the 
equation  
\begin{equation}  
\frac{d^2f}{d\tau^2} = \gamma\left(1 -  
\frac{I}{I_c}\right)\frac{1}{f^3} \,  \label{variational} 
\end{equation} 
where $\gamma = 4(\ln 4 + 1)/(27\zeta(3)) \simeq 0.29$, $I(\tau) =  
f^2(\tau)|F(\tau)|^2 = f^2(0)|F(0)|^2 = I_0$,  
and $I_c = (2\ln2 + 1)/(4\ln2 - 1)\simeq 1.35$.\footnote{Here 
$\zeta(p)$ is the Riemann zeta function.}    
Obviously, stationary solutions exist when the pulse power satisfies  
$I_0 = I_c$.   
Furthermore, the full solution of Eq.\ (\ref{variational}) is  
\begin{equation} 
f(\tau) =  
f(0)\sqrt{1 + \frac{\gamma}{f^2(0)}\left(1 - 
\frac{I_0}{I_c}\right)\tau^2} \ ,  
\end{equation} 
which shows that the stationary solution is unstable and either 
collapses to zero width in a   
finite time when $I_0 > I_c$, or diffracts monotonously towards 
infinite width when $I_0 < I_c$.

%%%%%%%%%%%%%%%%%%%%%%%%%%%%%%%%%%%%%%%% 
\section{Discussion and conclusion} 
%%%%%%%%%%%%%%%%%%%%%%%%%%%%%%%%%%%%%%%% 
 
From Sec.\ \ref{sec:two}, it is clear that the most interesting 
alternative is $I_{0}>I_{c}$, in which case  
the QED vacuum nonlinearities play a crucial role, leading to a 
collapse. In  
dimensional units the inequality $I_{0}>I_{c}$ roughly leads to 
$E_{{\rm init}}^{2}k^{2}r_{{\rm init}}^{2}>E_{{\rm crit}}^{2}$,  
where $E_{{\rm init}}$ and  
$r_{{\rm init}}$ are the initial electric field and beam radius, 
respectively, and the critical QED electric field is defined by 
$E_{{\rm crit}}^{2}=\varepsilon _{0}/\xi $. Naturally this unbounded 
self-focusing will  
eventually be saturated by some kind of higher order nonlinear 
mechanism.  
However, this will not occur before an electric field level of order 
$E \sim E_{{\rm crit}}$ is reached, which is the field strength when 
both our  
perturbative nonlinear calculation scheme and our starting expression,  
the  
Euler--Heisenberg Lagrangian, breaks down. For such extreme energy 
densities,  
higher order Feynman diagrams must be included in the QED description,  
and possibly the corresponding physical  
effects may then counteract the collapse scenario, resulting in a 
saturated  
beam radius $r_{{\rm sat}}\sim r_{{\rm init}}E_{{\rm init}}/E_{{\rm 
crit}}$. The possibility to  
reach such extreme intensity levels is of very much interest. 
However, it is clear that setting up the conditions necessary for vacuum 
light bullets to be formed, is a technological challenge. 
 
Presently the electric fields that can be supported by the walls before 
field emission takes place are of the order $10^8$ V/m \cite{Graber}. 
For such field strengths we note from the 
above estimates that we must  have $k r_{{\rm init}} \sim 10^{11}$, in
which case either  
the initial beam radius becomes unrealistically large (of the order of
km), or the  
wavelength to short (i.e. in the short uv or x-ray regime) for the
walls to be  
conducting. Two things should be stressed, however. Firstly  it should  
be noted that the experimentally possible field level before field
emission takes  
place have increased significantly in recent years
\cite{Graber}. Secondly,   
in our geometry there is no normal component of the electric field at
the wall  
surfaces, which may allow for a higher (central) field strength than 
normal, relaxing the necessary value on $k r_{{\rm init}} $.  A
slightly different but  
related technological question is whether the huge beam powers needed
for vacuum  
self-focusing can be reached.  This issue was discussed in Ref.\
\cite{segev},  
where it was concluded that it may   
occur within the next 10--15 years, given the current rate of 
technological improvement. 
 
A very interesting question from a principal point of view, would be 
whether  
it is possible to have fully three-dimensional QED-structures which do 
not  
require any guiding support. However, it seems that this issue cannot 
be  
addressed within a perturbational approach, and further research 
is thus required. 
 
In this paper we have shown that the photon--photon scattering, due  
to the exchange of virtual electron--positron pairs, that effectively  
gives rise to self-interaction terms in Maxwell's equations, can cause 
optical collapse in vacuum if the right conditions are met. More 
specifically, high intensity electromagnetic waves guided by two 
parallel conducting planes can form light bullets, which can collapse 
if the intensity of the beams are high enough.     
It is possible that higher order corrections to the 
Euler--Heisenberg Lagrangian can counteract such a collapse scenario, 
thus leading to stable optical vacuum structures. 
Further research is also necessary to find out more about how the
optimum configurations can be changed, both in the present paper and
in previous work (e.g.\ \cite{segev}).

%%%%%%%%%%%%%%%%%%%%%%%%%%%%%% 
 
%%%%%%%%%%%%%%%%%%%%%%%%%%%% 
\end{document}